\documentstyle[12pt]{article}
\date{}
\begin{document}
\title{{\LARGE\sf Metastable States in Spin Glasses and Disordered Ferromagnets
}}  
\author{
{\bf C. M. Newman}\thanks{Partially supported by the 
National Science Foundation under grant DMS-98-02310.}\\
{\small \tt newman\,@\,cims.nyu.edu}\\
{\small \sl Courant Institute of Mathematical Sciences}\\
{\small \sl New York University}\\
{\small \sl New York, NY 10012, USA}
\and
{\bf D. L. Stein}\thanks{Partially supported by the 
National Science Foundation under grant DMS-98-02153.}\\
{\small \tt dls\,@\,physics.arizona.edu}\\
{\small \sl Depts.\ of Physics and Mathematics}\\
{\small \sl University of Arizona}\\
{\small \sl Tucson, AZ 85721, USA}
}
\maketitle
\begin{abstract}
We study analytically $M$-spin-flip stable states in disordered short-ranged
Ising models (spin glasses and ferromagnets) in all dimensions and for all $M$.
Our approach is primarily dynamical and is based on the convergence of
$\sigma^t$, a zero-temperature dynamical process with flips of lattice animals
up to size $M$ and starting from a deep quench,
to a metastable limit $\sigma^\infty$. The results (rigorous and nonrigorous,
in infinite and finite volumes) concern many aspects of metastable states:
their numbers, basins of attraction, energy densities, overlaps, remanent
magnetizations and relations to thermodynamic states. For example, we show
that their overlap distribution is a delta-function at zero.
We also define a dynamics for 
$M=\infty$, which provides a potential tool for investigating ground state structure.
\end{abstract}
\small
\normalsize

\section{Introduction}
\label{sec:intro}

\subsection{Overview}
\label{subsec:overview}

Studies of spin glass dynamics often start from the assumption that their
anomalous, and still poorly understood, features arise from the presence of
a large number of ``metastable'' (i.e., locally stable) states within the
spin glass phase (many reviews are available; see, for example,
Refs.~\cite{Ryan,FHbook,BY}).  Although there exists plentiful (though
mostly indirect) evidence for the presence of many metastable states in
spin glasses, little hard knowledge of their properties has been obtained.
Most treatments of spin glass dynamics must therefore rely on assumptions
--- that often differ widely --- about their number, nature, and structure
\cite{aging1,aging2,aging3,Orbach1,aging4,aging5,aging6,aging7,aging8,aging9,aging10,Orbach2,Orbach3}.
Questions regarding metastability (and the accompanying ``broken
ergodicity'' \cite{Palmer82}) are important also in the study of other
disordered systems, such as glasses \cite{EAN,BMRS}, and of certain neural
network models \cite{Hopfield,AGS1,AGS2}.  Any information on spin glass
metastable states, obtained from first principles and without assumptions,
would therefore be highly useful.  (The reader who wishes to cut to the chase
is referred to Subsecs.~\ref{subsubsec:questions}, \ref{subsubsec:main} below, where
our results, providing such information, are summarized.)

Numerical simulations have provided much of the evidence for the existence
of metastability in spin glasses; indeed, the presence of metastability has
often been an impediment to studies of equilibrium properties
\cite{DMH,Heidelberg}, and has in turn led to new numerical techniques such as
simulated annealing \cite{CZ,KGV}.  Experiments are frequently interpreted
through the use of metastable states and are used to try to extract
information about them; early examples include ac susceptibility,
time-dependent magnetization, spin echo, M\"ossbauer effect, and others
\cite{FHbook}.  More recent experiments that may provide information on
metastable states include measurements of noise in mesoscopic spin glasses
\cite{Weissman} and aging
\cite{aging1,aging2,aging3,Orbach1,aging4,aging5,aging6,aging7,aging8,aging9,aging10,Orbach2,Orbach3}.
However, because assumptions about the number and structure of metastable
states must invariably be made, our general understanding of the role
played by metastability in spin glass dynamics remains relatively
primitive.
 
Because this understanding cannot be obtained through conventional
statistical mechanical tools, few analytical results are available, and are
usually confined to the case of 1-spin-flip (energetically) stable states.
In early work, Tanaka and Edwards \cite{TK}, Bray and Moore \cite{BM80},
and De~Dominicis {\em et al.\/} \cite{DGGO} studied their number in the
Sherrington-Kirkpatrick (SK) \cite{SK} (or equivalently, the
Thouless-Anderson-Palmer (TAP) \cite{TAP}) mean field spin glass.  They
found that the number of 1-spin-flip stable states in a system of $N$ spins
scaled as $\exp(0.1992N)$.  Nemoto \cite{Nemoto} studied the same set of
metastable states, and asserted both that their energy levels behaved as in
a random energy model, and that the barrier energy between them is an
increasing function of their Hamming distance.  Vertechi and Virasoro
\cite{VV}, in both analytical and numerical work, and confining their
analysis to the lowest energy (metastable) states, found results consistent
with the hypothesis that the energy barriers between metastable
states scale with their Hamming distance; they suggested that this
correspondence might explain the mean field ultrametric organization of
states.  Other work has also been done on the distribution of barriers in
the SK model \cite{IS}, as well as on metastable states in other mean-field
models, including the infinite-ranged $p$-spin-interaction spin glass
\cite{OF}, the spherical $p$-spin model \cite{FP}, the infinite-ranged
Potts glass \cite{Potts}, and related systems such as Kauffman's $N-k$
model \cite{Kauffman}.

There exist few theoretical results on metastable states in short-ranged
disordered systems in two or more dimensions even though results on these
would be important in interpreting laboratory experiments.  Rare analytical
results have been obtained on a one-dimensional spin chain with a
continuous coupling distribution symmetric about zero \cite{Li,EM,DG}.  It
was found that the number of 1-spin-flip stable states increases
exponentially with the system size (in \cite{Li}, metastable states of
greater than single spin stability were also examined).  Derrida and
Gardner \cite{DG} further showed that there existed a maximum magnetization
above which there existed no metastable states.  Bray and Moore \cite{BM81}
have presented a replica formalism for studying 1-spin-flip stable states
in finite dimensional spin glasses; using this formalism, they carried out
a stability analysis about mean field theory and studied some of the
properties of metastable states with higher energies.  More recently,
numerical studies \cite{RO} of the two-dimensional $\pm J$ spin glass seem
to indicate that as system size increases, the energy densities of the
(1-spin-flip) metastable states converge to a single value.

Summarizing, it appears that until now it has been difficult to obtain hard
analytical results on metastable states in short-ranged spin glasses in
dimension greater than one.  Aside from demonstrating that such states
almost certainly exist in spin glasses and are important in determining
their physics, neither experiment nor numerical work to date can provide
unambiguous and detailed information on their structure.  Both analytical
and numerical analyses that directly address the properties of metastable
states (as opposed to inferring their properties indirectly) have mostly
been confined either to mean-field or one-dimensional models, and are
usually limited to the study of $1$-spin-flip energetically stable states.

In this paper we provide rigorous results on metastable states that rely on
no approximations or assumptions.  We will analyze the properties of
metastable states in disordered spin systems (in particular, spin glasses
and random ferromagnets, both with continuous coupling distributions) in
all finite dimensions, and we will study states that are energetically
stable \cite{energy} up to a flip of any $M$ spins, where $M<\infty$ can be
arbitrarily large.  Both infinite volume and finite volume systems will be
addressed.

Before proceeding, we wish to add one cautionary note.  Although we believe
that the concept of metastable states is both interesting and useful in
understanding spin glass (and other) dynamics, we believe also that
alternative (but not necessarily orthogonal) formulations exist that have
the potential to provide this understanding without direct invocation of
such states.  These are fully real-space pictures, such as droplet-scaling
\cite{Mac,FH86,BM87} but possibly also others, that interpret
nonequilibrium spin glass dynamics following a quench through the resulting
domain structure \cite{FH88,KH}.  Such approaches have several advantages,
in our opinion, over those invoking metastable states (especially over
those that make no contact with real-space structure).  First, they require
fewer assumptions (most of which, however, remain neither verified nor
disproved) and those assumptions are typically more accessible to numerical
or analytical tests than those regarding metastable states.  Second, the
idea that the sample breaks up into domains, of whatever ultimate nature,
following a deep quench is appealing and likely correct.

While the distinction between the thermodynamic pure states and metastable
states of a system remains important, the overwhelming focus on metastable
states (divorced from real-space considerations) has led in part to the
common viewpoint that pure state structure is irrelevant to dynamics,
because the system is believed to spend all its time in a single pure
state.  We have shown elsewhere \cite{NS99a} that this is in general {\it
not\/} correct, particularly for nonequilibrium dynamics following a deep
quench.  The pure states are indeed relevant to dynamics, and at some level
metastability and metastable states (at both zero and positive temperature)
should be related to a description based upon the pure state structure.  We
will not discuss such a relation further in this paper, and will treat
metastable states independently from these considerations.  If the above
caveat is kept in mind, then the study of metastable states can provide a
useful (but not orthogonal) complement to real-space approaches based on
thermodynamic pure state structure.

\subsection{Summary of results.}
\label{subsec:summary}

Although most of our results will apply to many types of disordered
systems, we consider for specificity the Ising spin Hamiltonian on the
$d$-dimensional infinite cubic lattice $Z^d$,
\begin{equation}
\label{eq:Hamiltonian}
{\cal H}=-\sum_{<xy>} J_{xy}\sigma_x\sigma_y\ .
\end{equation}
Here the sites $x$ are in 
$Z^d$, the spins $\sigma_x=\pm 1$, and the sum is over
nearest neighbors.  The couplings $J_{xy}$ will be taken to be independent,
identically distributed random variables (though occasionally we will
examine other cases); we require of their 
common distribution that it be
continuous and have finite mean (and, for some of our results, further
requirements).  We denote by ${\cal J}$ a particular
realization of all the couplings.  

Both the spin glass and ferromagnetic cases will be considered.  In the
first case, the couplings can take on either positive or negative values,
typically (but not necessarily) symmetrically distributed about zero; this
is the Edwards-Anderson (EA) Ising spin glass model \cite{EA}.  In the
second case, the couplings take on only positive values.  A Gaussian
distribution of couplings with zero mean is most commonly used in the spin
glass case, while a uniform distribution of couplings in the interval
$[0,J]$ typifies the random ferromagnet.  While our results are not
restricted to these distributions, we will use them often throughout the
paper for clarity.

A 1-spin-flip stable state is defined as an infinite-volume spin
configuration whose energy as given by Eq.~(\ref{eq:Hamiltonian}) cannot be
lowered by the flip of any single spin.  Similarly, an $M$-spin-flip stable
state ($M<\infty$) is an infinite-volume spin configuration whose energy
cannot be lowered by the flip of any subset of $1,2,\ldots,M$ spins.
Finally, a ground state is an infinite-volume spin configuration whose
energy cannot be lowered by the flip of {\it any\/} finite subset of spins.

All of the above definitions can be extended in a natural
way to finite-volume metastable states with specified boundary condition.
For finite-volume ground states, however, we can use the alternative (and
more natural) definition that it is the spin configuration (or spin
configuration pair, in the case of spin-flip-symmetric boundary conditions,
such as free or periodic) that has the lowest energy given the specified
boundary condition.  It is easily seen both that the definition given in
the preceding paragraph is equivalent to this in finite volumes, and that
the second definition has no natural extension to infinite volumes.

It has occasionally been noted that a definition of the energy (or free
energy) barrier confining a metastable state remains ambiguous at least
until a specific dynamics is defined.  We note here that this problem
does not exist for the definition of (energetically) metastable states
themselves, which can be defined solely through the use of a Hamiltonian
such as Eq.~(\ref{eq:Hamiltonian}).  Nevertheless, the essential approach
of this paper will be to study the metastable states by using dynamics to obtain
a natural ensemble of these states. 

\subsubsection{Questions.}
\label{subsubsec:questions}

Given these definitions, we can now ask for information about the
metastable states of disordered systems such as spin glasses.  We will not
attempt to be precise here, and some concepts (e.g., basin of attraction)
remain to be defined.  This and the following subsection are intended only
to serve as an overview of our main results, and as a reference point when
reading later sections of the paper.

\medskip

The most basic questions about metastability include:

\medskip

\noindent 1) At the most basic level, can the existence of metastable
states be proved?  If yes, how many 1-spin-flip, 2-spin-flip, $\ldots$
metastable states exist in $d$ dimensions?  Does the number of
$M$-spin-flip stable states vary with $M$ or $d$?  If this number is
infinite for some $M$ and/or $d$, is it a countable or uncountable
infinity?

\noindent 2) Given an initial spin configuration $\sigma^0$ (following a
deep quench) and a specified zero-temperature dynamics (such as ordinary
Glauber dynamics), does $\sigma^t$, at time $t$, evolve towards a single
final metastable state $\sigma^\infty$ (i.e., do the dynamics converge)?
If so, how much of the initial information contained in the starting spin
configuration is contained in the final state, and how much varies with the
particular realization of the dynamics (nature vs.~nurture)?

\noindent 3) How large are the basins of attraction of the metastable
states?

\noindent 4) What is the distribution of energy densities of the metastable
states?

\noindent 5) What is the metastable state structure in configuration space?
For example, does there exist any nontrivial overlap distribution, in
finite or infinite volume?  Is there any scaling of the barrier height
(defined in some suitable or reasonable way) between 1-(or
higher)-spin-flip stable states with their Hamming distance, as has
sometimes been claimed?

\noindent 6) Does the number and structure of the various types of
metastable states differ for those that arise dynamically from two
independent starting configurations, as opposed to those that evolve from
the same initial configuration?  (This is somewhat different from the
questions asked in 2), though not orthogonal.)

\noindent 7) What does 1) imply about how the number of metastable states
scales with volume in finite samples?  Do the answers to 2)--6) change for
(large) finite volumes?

\noindent 8) What is the remanent magnetization in $d$-dimensional spin
glasses when the initial spin configuration is uniformly $+1$?

\noindent 9) Is there any correspondence between pure and metastable
states?  More precisely, is the spin configuration corresponding to a
typical metastable state in the domain of attraction of a single pure state
(at positive temperature, assuming multiple pure states) or ground state
(at zero temperature)?

\noindent 10) Do the answers to these questions about metastable states
provide any interesting thermodynamic information, such as the structure of
ground states at zero temperature or pure states at positive temperature?

\subsubsection{Main results.}
\label{subsubsec:main}

In this subsection we present the ``short'' answers to the above questions,
without discussion or elaboration.  A fuller discussion, without which
these answers should be regarded as sketchy and incomplete, will be
provided in later sections.

The numbers refer to the corresponding questions from the previous
subsection.  The section of the paper in which the claim made below is
proved and/or discussed is also given.

\medskip

\noindent 1) In an infinite system, the Hamiltonian
Eq.~(\ref{eq:Hamiltonian}) displays {\it uncountably\/} many $M$-spin-flip
stable states, for all finite $M\ge 1$ and for all finite $d\ge 1$
(Sec.~\ref{sec:overlaps}).

\noindent 2) For almost every ${\cal J}$, $\sigma^0$ and dynamics
realization $\omega$ (to be defined in Sec.~\ref{sec:dynamics})
\cite{almost}, a final
state $\sigma^\infty$, depending on the particular dynamics, exists.
Put another way, every spin flips only finitely many times
(Sec.~\ref{sec:convergence}) .  (This result is not obvious and indeed is
not the case for other systems, such as homogeneous ferromagnets on $Z^d$ --- at
least for low $d$;
see, e.g., \cite{NS99a,NS99b,NNS}.)  In the usual 1-spin-flip Glauber
dynamics in $1D$, precisely half the spins in $\sigma^\infty$ are
completely determined by $\sigma^0$, with the other half completely
undetermined by $\sigma^0$.  For higher $d$ and the same dynamics, it can
be shown that a dynamical order parameter $q_D$, measuring the
percentage dependence of $\sigma^\infty$ on $\sigma^0$,
is strictly between 0 and 1
(Sec.~\ref{subsec:nvn}).  (All results hold for almost every ${\cal J}$,
$\sigma^0$, and $\omega$.)

\noindent 3) The basins of attraction of the individual metastable states
are of negligible size.  That is, almost every initial configuration
$\sigma^0$ is on a boundary between (two or more) metastable states
(Sec.~\ref{sec:boa}).  Equivalently, the union of the domains of attraction
of {\it all\/} of the metastable states forms a set of measure zero (in the
space of all $\sigma^0$'s).  (A similar result for pure states was proved
in \cite{NS99a}.)

\noindent 4) For any $k$, almost every $k$-spin-flip stable state has the
same energy density, $E_k$.  Moreover, the dynamics can be chosen so that
$E_1>E_2>E_3>\ldots$ , and furthermore $E_k$ for any finite $k$ is larger
than the ground state energy density, which of course is the limit of $E_k$
as $k \to \infty$ (Sec.~\ref{sec:energies}).

\noindent 5) Almost every pair of metastable states (either two
$k$-spin-flip stable states or one $k$- and one $k'$-spin-flip stable
state) has zero spin overlap.  This conclusion does not change when one
restricts attention to any (positive measure) subset of metastable states
(Sec.~\ref{sec:overlaps}).

\noindent 6) For two metastable states arising from two independently
chosen starting configurations $\sigma^0$ and $\sigma'^0$, the answers
given above hold.  For almost any pair of metastable states arising from
the {\it same\/} $\sigma^0$, the answers in 1), 3), and 4) still hold, but
the answer to 5) is modified: it remains true that almost every pair has
the {\it same\/} overlap, but the overlap is now positive, and equal to the
quantity $q_D$ (Sec.~\ref{sec:samesigma}).

\noindent 7) The number of metastable states in finite samples scales (for
sufficiently large volumes) exponentially with the volume in general $d$
for states of any stability.  It is already known that the number of
1-spin-flip stable states in a one-dimensional chain of length $L$
increases as $2^{L/3}$ \cite{Li,DG}.  Exact results can be obtained in
higher dimensions for some other models.  In (large) finite volumes, the
answers to 2) and 4)-6) still hold (but there will be some smearing of the
delta-functions due to finite-volume effects).  For 3), the size of the
basins of attraction of the metastable states falls to zero as volume
increases (Sec.~\ref{sec:finitevolume}).

\noindent 8) The remanent magnetization in one dimension is known to be
$1/3$ \cite{FM}. In higher dimensions, a heuristic calculation suggests a
lower bound on the remanent magnetization that for large $d$ behaves like
$e^{-2d\log(d)}$ (for a Gaussian spin glass).  Exact results can be
obtained in all $d$ for some other models, to be discussed in
Sec.~\ref{sec:remanence}.

\noindent 9) At zero temperature, almost no metastable
state 
should be ``contained'' within a single ground state.  If more than one pure
state exists at some positive temperature, then almost no metastable state
should be contained within 
a single pure state.  That is, almost every metastable
state should be on a ``boundary'' in 
configuration space between multiple pure or
ground states (Sec.~\ref{sec:boa}).

\noindent 10) Information on metastable states so far does not seem to
provide information on infinite-volume pure or ground states.  That is, we
will see that one can have a situation (the $2D$ disordered ferromagnet)
where there exists an uncountable number of (infinite-volume) $M$-spin-flip
stable states for $M$ arbitrarily large, but in which there exists only a
single pair of pure states at low temperature (Sec.~\ref{sec:ground}).  In
situations of this kind, the presence of many metastable states could conceivably lead to
difficulties in interpreting numerical studies of equilibrium properties,
such as the number of pure (or ground) states.

\medskip

The claims made in 1) -- 6) will be proven rigorously.  For 7), the
claim of exponential scaling of the number of states will be proven
rigorously; the claim concerning overlaps in finite volumes will be proven
rigorously when $M=1$ for a class of disordered systems thermodynamically
equivalent to ordinary spin glasses and random ferromagnets.  This result
should hold also for $M>1$ and for ordinary spin glasses and random
ferromagnets; for these, heuristic arguments will be presented.  The claims
of 8) include rigorous exact results for certain models and heuristic lower
bounds for other disordered systems. The claims of 9) are motivated
by 3), but have not yet been formulated in a rigorous way.
The claim of 10) is based on a
conjecture that is widely believed but that remains to be proven
rigorously.

\subsubsection{Outline of rest of paper.}
\label{subsubsec:outline}

In Sec.~\ref{sec:dynamics}, we present the dynamical processes to be
considered.  The single-spin dynamics is simply the zero-temperature limit
of the usual Glauber dynamics, but we present also a multi-spin-flip
dynamics.  Sec.~\ref{sec:convergence} presents arguments showing
convergence of the dynamics to final states in several contexts: both
finite- and infinite-volume ordinary spin glasses and random ferromagnets
(hereafter referred to simply as ordinary disordered models) in any
dimension, for strongly and highly disordered models, and finally for
certain types of 
homogeneous systems.  In Sec.~\ref{sec:overlaps}, we prove
that the number of $M$-spin-flip stable states is uncountably infinite for
ordinary disordered models, in any dimension and for any $M$, and that the
spin overlap distribution is a delta-function at zero.  We also discuss
there implications for arguments that barriers between metastable states
scale with their Hamming distance.  In Sec.~\ref{sec:energies} we show that
the energy densities of all $M$-spin-flip stable states (except for a set
of measure zero) are the same, and show that a natural choice of dynamics
leads to lower energy densities for states of higher stability.  We also
present there a cautionary discussion about how to interpret and use these
and related conclusions.

We then present, in Sec.~\ref{sec:remanence}, a calculation of the
``remanent overlap'' (and for spin glasses, remanent magnetization, which
is a special case) for highly disordered models, and also provide a
(nonrigorous) lower bound of this quantity for ordinary disordered models
in general dimensions.  In the same section we also compute the dynamical
order parameter $q_D$ for
ordinary disordered models in one dimension and highly disordered models in general
dimensions.  In Sec.~\ref{sec:boa}, we show that the basins of attraction
of almost every metastable state have measure zero, and remark that no
metastable states (as always, aside from a set of measure zero) should
themselves lie completely in the basin of attraction of any ground (or at
positive temperature, pure) state.  In Sec.~\ref{sec:samesigma} we show
that the spin overlap distribution for two metastable states dynamically
evolved from the {\it same\/} (random) starting configuration is $q_D$,
almost surely.  In Sec.~\ref{sec:finitevolume} we re-examine many of the
above results for finite-volume disordered systems, and show that their
qualitative features persist in large finite volumes, and that quantitative
agreement with the infinite-volume results is increasingly better as the
volume increases.  In Sec.~\ref{sec:ground}, we present a dynamics that
generates infinite-volume {\it ground\/} states, and discuss their relation
with metastable states.  Finally, in Sec.~\ref{sec:conclusions}, we present
our conclusions.

\section{Dynamics}
\label{sec:dynamics}

Theoretical studies of metastable states usually look directly for
1-spin-flip stable configurations for the Hamiltonian (as in, e.g.,
\cite{Li,EM,DG}) or for 1-spin-flip stable solutions of self-consistent
equations for the magnetization (as in, e.g., \cite{BM80}).  Here we
propose instead a {\it dynamical\/} approach, in which the time evolution
of the system is exploited as a theoretical tool in determining the answers
to the questions posed in Sec.~\ref{subsec:summary}.  We start by
describing the dynamics that will be used.

We begin by considering the standard zero-temperature Glauber
single-spin-flip dynamics.  In every dynamical process considered in this
paper, the coupling realization ${\cal J}$ is taken to be fixed.  We denote
by $\sigma^0$ the initial (time zero) infinite-volume spin configuration on
$Z^d$.  The starting state $\sigma^0$ is chosen from the
(infinite-temperature) ensemble in which each spin is equally likely to be
$+1$ or $-1$, independently of the others.  The spin configuration is
updated asynchronously, in that a single spin at a time is chosen at
random, and then always flips if the resulting configuration has lower
energy and never flips if the resulting configuration has higher energy.
(Because the coupling distribution is continuous, there is no possibility
of a flip costing zero energy.  In models where zero-energy flips can
occur, as in uniform ferromagnets \cite{NS99a,NS99b,NNS} or $\pm J$ spin
glasses \cite{GNS}, the standard rule is that the chosen spin then flips
with probability 1/2.)

The notion of choosing a spin ``at random'' needs clarification for an
infinite-volume system.  More precisely, the (continuous time) dynamics is
given by independent (rate 1) Poisson processes at each $x$ corresponding
to those times $t$ at which the spin at $x$ looks at its neighbors and
determines whether to flip.  We denote by $\omega_1$ a given realization of
this zero-temperature single-spin flip dynamics; so a given realization
$\omega_1$ would then consist of a collection of random times $t_{x,i}$
($x\in Z^d$, $i=1,2,\ldots$) at every $x$ when spin flips for the spin
$\sigma_x$ are considered.

Given the Hamiltonian (\ref{eq:Hamiltonian}) and a specific ${\cal J}$,
$\sigma^0$, and $\omega_1$, a system will evolve towards a single
well-defined spin configuration $\sigma^t$ at time $t$.  It is important to
note that these three realizations (coupling, initial spin, and dynamics)
are chosen independently of one another.  The continuous coupling
distribution and zero-temperature dynamics together guarantee that the
energy per spin $E(t)$ is always a monotonically decreasing function of
time.

The above dynamics is commonly used in a variety of problems.  We now
introduce a dynamics that employs multiple-spin flips.  Consider a dynamics
in which rigid flips of all lattice animals (i.e., finite connected subsets
of $Z^d$, not necessarily containing the origin) up to size $M$ spins can
occur.  One could restrict flips to only simply connected lattice animals
(i.e., no holes), but we will not do so.  The case $M=1$ is the single-spin
flip case just described; $M=2$ corresponds to the case where both
single-spin flips and rigid flips of all nearest-neighbor pairs of spins
are allowed; and the case of general $M$ corresponds to flips of 1-spin,
2-spin, 3-spin, $\ldots$ $M$-spin connected clusters.  A specific
realization of this $M$-spin-flip dynamics will be denoted $\omega_M$.

The probability measure $P_M$ from which a dynamical realization $\omega_M$
is taken must be chosen so that the resulting dynamics is sensible, i.e.,
so that the dynamics leads to a single, well-defined $\sigma^t$ for almost
every ${\cal J}$, $\sigma^0$, and $\omega_M$.  Furthermore, we wish the
dynamics to remain sensible even in the limit $M\to\infty$.  An initial
requirement on $P_M$ is that the probability that any fixed spin considers
a flip in a unit time interval remains of order one, uniformly in $M$.
Such a choice would guarantee, for example, that the probability in
$P_M$ that a spin considers a flip in a time interval $\Delta t$
vanishes as $\Delta t\to 0$, uniformly in $M$.   
A further 
requirement for the dynamics to be well-defined is
that information not propagate arbitrarily fast throughout the
lattice as $M$ becomes arbitrarily large.

We therefore construct our dynamics as follows: $P_M$ for $M$ fixed
assigns all simply connected lattice animals of size $k$ (i.e., containing
$k$ spins) a dynamics chosen from a Poisson process as in the
single-spin-flip case, but with rate $R_k>0$ depending on $k$, for
$k=1,2,\ldots,M$.  We take, as before, $R_1=1$, and in general will require
that $R_{k+1}<R_k$ for all $k$.  As always, the dynamical process governing
the flipping of any lattice animal is independent of that for all others.

It is not hard to show that for any spin to flip at a rate of order one,
independent 
of $M$ in the multi-spin-flip dynamics, it is enough to require
that $\sum_{k=1}^\infty h_kR_k<\infty$, where $h_k$ is the number of
lattice animals of size $k$ that contain the origin.  This number scales
exponentially in $k$, with the constant in
the exponential (generally not known for most $d$) dependent
on the lattice type and dimensionality \cite{SA}. 
We therefore define our dynamics so that $R_k\sim\exp{[-a(d)k]}$, where
$a(d)>0$ depends only on dimension.  In order that information not propagate
infinitely fast even after $M\to\infty$, we choose $a(d)$ large enough
so that $h_kR_k$ still decays exponentially fast as $k\to\infty$ (see also
Theorem~3.9 in Chap.~I of \cite{Liggett}).
There might exist slower falloffs of
$R_k$ with $k$ that would also give a reasonable dynamics, but our purpose
here is only to point out that such dynamics do exist.

We emphasize that we are not proposing this multi-spin-flip dynamics in
order to model dynamical processes in actual spin glasses (although it
could conceivably be useful for that purpose).  Its intended use is rather
as a theoretical tool to help elucidate the structure of metastable states.
We now proceed to show how this may be done.

\section{Convergence of the Dynamics}
\label{sec:convergence}

In this section, we study the question of convergence of 
$\sigma^t$ to a
final (metastable) state $\sigma^\infty$.  
As always, we consider a disordered Ising spin
system with energy given by Eq.~(\ref{eq:Hamiltonian}), whose coupling
realization ${\cal J}$ is fixed throughout the dynamical process.  Unless
otherwise specified, ${\cal J}$ will be chosen from a continuous coupling
distribution with finite mean (but other distributions will also be briefly
discussed).  The initial spin configuration $\sigma^0$ is chosen from the
(infinite-temperature) distribution described at the beginning of
Sect.~\ref{sec:dynamics}.  Strictly speaking, the dynamical process
corresponds to that following an instantaneous quench from infinite to zero
temperature.  Physically, such a process is often used to model the
behavior of systems following a deep quench from high to low temperature.

We will consider the system's evolution to a final state in both the
finite-volume and infinite-volume cases.  The question of convergence is
not so obvious in the 
infinite-volume case, but is rather easy in the finite-volume case,
so we will begin there.

\subsection{Finite volumes}
\label{subsec:finiteconvergence}

We will denote by $\Lambda_L\subset Z^d$ the $L^d$ cube centered at the
origin and by $|\Lambda_L|$ the number of sites in $\Lambda_L$.  
Given some specified boundary condition (periodic, fixed, free, etc.)
on $\partial\Lambda_L$, the boundary of $\Lambda_L$, there
is a unique (with respect to spin configurations, modulo a global spin flip
if the boundary condition is spin-symmetric) minimum $E_{\rm min}^{(L)}$
over all spin configurations of the energy within $\Lambda_L$.  The
uniqueness, for almost every ${\cal J}$, is a consequence of the coupling
distribution being continuous. Similarly, for almost every ${\cal J}$ there
will be a minimum energy change $\Delta_{\rm min}^{(L)} > 0$ over all possible
flips (of lattice animals strictly contained in $\Lambda_L$ up to size 
$M < |\Lambda_L|$) in all of the
$2^{|\Lambda_L|}$ spin configurations in $\Lambda_L$.  The actual value of
$\Delta_{\rm min}^{(L)}$ will depend on ${\cal J}$, the boundary condition,
and the choice of dynamics, i.e., the value of $M$ in $P_M$.  The energy
$E^{(L)}(0)$ at time 0 is finite, so the total number of spin flips is
bounded from above by $(E^{(L)}(0)-E_{\rm min}^{(L)})/\Delta_{\rm
min}^{(L)}$.  It follows that the spin configuration converges, after a
finite number of spin flips in finite time, to some limiting
$\sigma_{(L)}^\infty$ (depending on $\sigma_{(L)}^0$ and $\omega_M$).  We
now turn to the more interesting case of dynamical convergence to a
limiting spin configuration in infinite volumes.

\subsection{Infinite volumes}
\label{subsec:infiniteconvergence}

\subsubsection{``Ordinary'' spin glasses and random ferromagnets}
\label{subsubsec:ordinary}

Given the Hamiltonian (\ref{eq:Hamiltonian}) and a continuous coupling
distribution with finite mean, it was proved in \cite{NNS} for $M=1$ that
every spin flips only finitely many times for almost every ${\cal J}$,
$\sigma_0$, and $\omega_1$.  This was implied by a more general result that
even if the coupling distribution is not continuous, (in almost every
realization) there can be only finitely many energy-decreasing flips (as
opposed to zero-energy flips) of any spin.  Given both the dynamics and the
continuity of the coupling distribution, every spin flip strictly decreases
the energy, and the implication follows.  We now sketch the proof given in
\cite{NNS}, modified very slightly to incorporate the more general
$M$-spin-flip dynamics; we refer the reader to \cite{NNS} for technical
details.

We denote by $\sigma_x^t$ the value of the spin at $x$ for fixed ${\cal
J}$, $\sigma^0$, and $\omega_M$.  Define
\begin{equation}
\label{eq:energy}
E(t)=-(1/2)\overline{\sum_{y:||x-y||=1}J_{xy}\sigma_x^t\sigma_y^t}
\end{equation}
where the overbar indicates an average over $({\cal J}$, $\sigma^0, 
\omega_M)$ and $||x-y||$ denotes Euclidean distance.  
By translation-ergodicity of the distributions from which
${\cal J}$, $\sigma^0$, and $\omega_M$ are chosen, and using the assumption
that the distribution of ${\cal J}$ has finite mean, it follows that $E(t)$
exists, is independent of $x$, and equals the energy density (i.e., the
spatial-average energy per site) at time $t$ in almost every realization of
${\cal J}$, $\sigma^0$, and $\omega$.

Clearly $E(0)=0$ (because of the spin-flip symmetry of the distribution of
$\sigma^0$) and $E(\infty)\ge -d\overline{|J_{xy}|}$.  We now choose any
fixed number $\epsilon>0$, and let $N_x^\epsilon$ be the number of flips
(over all time) of the spin at $x$ (i.e., of lattice animals containing
$x$) that lower the total energy by an amount $\epsilon$ or greater.
Because $-d\overline{|J_{xy}|}\le E(\infty)\le -(\epsilon / M)
\overline{N_x^\epsilon}$, it follows (for almost every ${\cal J}$,
$\sigma^0$, and $\omega_M$) that for every $x$ and every $\epsilon>0$,
$N_x^\epsilon$ is finite.  Then if $\epsilon_x$ is the minimum possible
magnitude of the energy change resulting from a flip of a lattice animal
containing $x$, we need only show that $\epsilon_x>0$ for every $x$.  The
value of $\epsilon_x$ of course varies with $x$ and will depend on both
${\cal J}$ and the value of $M$ in the dynamics measure $P_M$.  Let
$\Delta_{(k,x)}$ be the magnitude of the minimum energy change, in all spin
configurations, over flips of all lattice animals of size $k$ containing
$x$; clearly $\Delta_{(k,x)}>0$.  Then $\epsilon_x=\min_{1\le k \le
M}\Delta_{(k,x)}>0$ because $M<\infty$.

We have therefore proved for any $M < \infty$
the existence of a limiting state $\sigma^\infty$
for almost every ${\cal J}$, $\sigma^0$, and $\omega_M$.
The final state $\sigma^\infty$ of course depends on all three
realizations, and will be an $M$-spin-flip stable state.
Before exploring the consequences of this result, we turn briefly to 
a discussion of some other systems.

\subsubsection{Strongly and highly disordered models}
\label{subsubsec:hdm}

There is a class of ``strongly disordered'' coupling distributions, where
the mechanism for convergence of {\it single\/}-spin-flip dynamics is more
localized \cite{NNS} than the one given just above.  This class includes
distributions with infinite mean as well as ones with finite mean (although
we retain the requirement that the coupling distribution be continuous).
These are coupling distributions such that ``influence percolation''
\cite{NN} does not occur on ${\bf Z}^d$; we note that this requirement
yields a $d$-dependent class of distributions.  The reason for convergence
of dynamics is different in these cases, and a new approach based on the
idea of influence percolation is needed.  To discuss this we first describe
the notion of influence.

We say that the spin at $y$ can influence the spin at $x$ (where
$\|x-y\|=1$) if changing $\sigma_y$ can alter whether the energy
change resulting from a flip of $x$ 
is less than (or equal to, or greater than) zero in some spin
configuration.  So, for example, if the coupling $J_{xy}=0$ than $y$
cannot influence $x$ and vice-versa.  (This possibility, and also that of
zero energy changes, is excluded here, however, because we assume that the
coupling distribution is continuous.)  If $J_{xy} \ne 0$, then the
(necessary and sufficient) condition that $y$ can influence $x$ is
\cite{NN}
\begin{equation}
\label{eq:influence}
|J_{xy}| \ge |\sum_{z:\|x-z\|=1,z\ne y} \sigma'_zJ_{xz}|
\end{equation}
for some choice of the $\sigma'_z$'s (in $\{-1,+1\}$).  Because the
condition (\ref{eq:influence}) cares only about the coupling magnitudes and
not the signs, the discussion applies equally well to spin glasses and
random ferromagnets.

We now consider the graph consisting of all sites in ${\bf Z}^d$ but only
those bonds $\{x,y\}$ such that either $x$
can influence $y$ or $y$ can influence $x$ or both.  The properties of this
graph (called the {\it influence graph\/} in \cite{NN}) that are valid for
almost all ${\cal J}$, will depend on both $d$ and the coupling
distribution.  If there is no percolation of the influence graph (i.e., if
given some ${\cal J}$, all the clusters of the influence graph are finite)
and there is no possibility of zero energy flips, then every spin
$\sigma_x$ can flip at most finitely many times (for every $\sigma^0$ and
for almost every $\omega_1$).  This is because the dynamics is effectively
localized: the dynamics on ${\bf Z}^d$ of the infinite-volume spin
configuration breaks up into dynamics on disconnected finite regions.  The
result then follows (as in the analysis above of finite-volume dynamics).  
If influence percolation {\it does\/} occur, then no conclusions can be drawn
(without further information) on whether spins can flip infinitely often.
We note that for $d=1$, any continuous coupling distribution will result in
influence nonpercolation.

An example of a system where influence nonpercolation occurs (and so the
dynamics converges) is the ``highly disordered'' model of
\cite{NS94,BCM,NS96a}.  Here the couplings are volume-dependent and
``stretched out'' so that in large finite volumes, the magnitude of any
coupling is at least twice that of the next smaller one and no more than
half that of the next larger one.  However, influence nonpercolation can
also occur in less extreme situations, in particular the class of models we
call ``strongly disordered''.  Roughly speaking, these are models in which
the above condition on the stretching of the couplings typically holds up
to some maximum size volume (which still needs to be sufficiently large),
but not for arbitrarily large volumes.  For a more detailed description,
see \cite{NNS}.

\subsubsection{Other systems}
\label{subsubsec:other}

It is not difficult to see that the proof outlined in
Subsec.~\ref{subsubsec:ordinary} allows for a restatement of the dynamics
convergence theorem as follows: given $M$-spin-flip zero-temperature
dynamics in an infinite spin system where the energy per site is bounded,
and the initial spin configuration is chosen from a spatially ergodic
measure, there can (with probability one) be only finitely many flips that
cause a {\it nonzero\/} energy change.  We can therefore apply this result
not only to disordered systems with noncontinuous coupling distributions,
but also to homogeneous systems such as uniform ferromagnets or
antiferromagnets.  Here the theorem implies that the question of
convergence is lattice-dependent.  For example, every spin flip will be
energy-lowering in a uniform ferromagnet on a hexagonal (honeycomb) lattice
in two dimensions, so here too the dynamics will almost always converge
from a random initial spin configuration \cite{NS99b}.

What about uniform ferromagnets on square lattices?  Here we have proved
\cite{NS99a,NNS} that the opposite is true: for almost every $\sigma^0$ and
$\omega_1$ (the result easily extends to multi-spin-flip dynamics, but we
will not do so here), there is no convergence of the dynamics because every
spin flips {\it infinitely\/} many times.  It must remain true that every
spin undergoes only finitely many energy-lowering flips, so therefore every
spin must undergo infinitely many {\it zero\/}-energy flips.  A more global
viewpoint \cite{NS99a} is that there exists no finite time after which the
spins within some fixed, finite region remain in a single phase; that is,
domain walls forever sweep across the region.  We do not yet know what
happens in uniform ferromagnets on $Z^d$ in dimensions higher than two,
although numerical simulations \cite{Stauffer} indicate the
possibility of dynamical convergence in five and higher dimensions.

Finally, we briefly discuss the $\pm J$ spin glass (and related
models).  In two dimensions, we
can show \cite{GNS} that this is an intermediate case: (for almost every
$\sigma^0$ and $\omega_1$) a positive fraction of spins flip infinitely
many times and a positive fraction flip only finitely many times.  Similar
behavior occurs in spin glasses or random ferromagnets with other noncontinuous
distributions (e.g., the couplings can take on only two or a finite number
of values, and the distribution need not be symmetric about zero).  In all
of these, a limiting state $\sigma^\infty$ does not exist \cite{Jain}.
For noncontinuous distributions other than $\pm J$ models, these conclusions
remain valid for all $d \ge 2$ (whether that is so for $\pm J$ models
is unclear).

A discussion of these systems was included only for comparison purposes;
our primary interest in this paper will be in ordinary spin glasses and
random ferromagnets with continuous coupling distributions.  We now examine
the consequences of the results from this section.

\section{Numbers and Overlaps of Metastable States}
\label{sec:overlaps}

In Sec.~\ref{sec:convergence} we established that our $M$-spin-flip
dynamics converges to a final state $\sigma^\infty$ for almost every ${\cal
J}$, $\sigma^0$, and $\omega_M$.  We will hereafter denote by
$\sigma_M^\infty$ the final state reached in this way.  By the definition
of the measure $P_M$ from which the dynamical realizations $\omega_M$ are
chosen, it immediately follows that $\sigma_M^\infty$ is an $M$-spin-flip
stable state (for ${\cal J}$), which is a function also of $\sigma^0$ and
$\omega_M$.

It will be convenient to use a shorthand notation where (for fixed ${\cal
J}$) $\sigma^\infty_M$ denotes $\sigma^\infty_M(\sigma^0,\omega_M)$ and
$\sigma'^\infty_{M'}$ denotes $\sigma^\infty_{M'}(\sigma'^0,\omega'_{M'})$,
where $\sigma'^0$ and $\omega'_{M'}$ are chosen independently of $\sigma^0$
and $\omega_M$.  When $M'=M$, $\sigma^\infty_M$ and $\sigma'^\infty_{M}$
represent a pair of replicas. We define the overlap $Q_{M,M'}$ of
$\sigma^\infty_M$ and $\sigma'^\infty_{M'}$ in the usual way:

\begin{equation}
\label{eq:overlap}
Q_{M,M'}=Q({\cal J},\sigma^0,\omega_M,\sigma'^0,\omega'_{M'})=
\lim_{L\to\infty}|\Lambda_L|^{-1}\sum_{x\in\Lambda_L}\sigma_{M,x}^\infty 
\sigma'^{\infty}_{M',x}
,
\end{equation}
where $\sigma_{M,x}^\infty$ is the value of the spin $\sigma_x$ in the
metastable state $\sigma^\infty_M$ and $\sigma'^{\infty}_{M',x}$ is the value
of $\sigma_x$ in $\sigma'^\infty_{M'}$.
When $M'=M$, $Q_{M,M}$ is the overlap
of the replicas $\sigma^\infty_M$ and $\sigma'^\infty_{M}$.

We now show, in the following theorem, that for any finite $M$ there is an
uncountable infinity of $\sigma^\infty_M$'s \cite{uncountable}, and that
almost every pair $\sigma^\infty_M$, $\sigma'^\infty_{M}$
(as $\sigma^0, \omega_M, \sigma'^0, \omega'_{M'}$ vary
independently) has overlap zero.

\smallskip

{\it Theorem 1.\/} In a disordered spin system with
Hamiltonian~(\ref{eq:Hamiltonian}), for almost every fixed ${\cal J}$
chosen from a continuous coupling distribution with finite mean, there is
an uncountable infinity of $M$-spin-flip-stable states for any $M$
\cite{uncountable}.  Furthermore, for any $M$ and $M'$, almost every pair
has overlap zero; i.e., (for almost every ${\cal J}$) the infinite-volume
overlap distribution of $Q_{M,M'}$ is a single delta function at zero.

\smallskip

{\it Proof.\/} We first show that almost every pair of metastable states,
$(\sigma^\infty_M,\sigma'^\infty_{M'})$, has zero overlap, and then,
by taking $M'=M$, show how
this implies an uncountable infinity of metastable states.  For a fixed (finite)
$M$ and almost every ${\cal J}$, we showed in Sec.~\ref{sec:convergence}
that for almost every $\sigma^0$ and $\omega_M$ the dynamics converge to a
limiting metastable state $\sigma^\infty_M(\sigma^0,\omega_M)$.  Consider
two such final states $\sigma^\infty_M$ and $\sigma'^\infty_{M'}$, as defined
above.  Clearly their overlap $Q_{M,M'}$ is a measurable, translation-invariant
function of its five arguments.  Moreover, because each of the
five distributions from which ${\cal J}$, $\sigma^0, \omega_M,
\sigma'^0, \omega'_{M'}$ are chosen has the property of translation-ergodicity
(see \cite{NS97} for a discussion of this property and its use), it follows
that the same property holds for the joint (product) distribution of
$({\cal J},\sigma^0,\omega_M,\sigma'^0, \omega'_{M'})$.  The
translation-invariance of the random variable $Q$ (which is immediate from
the right-hand side of (\ref{eq:overlap})) then implies that it must be
constant for almost every realization of $({\cal
J},\sigma^0,\omega_M,\sigma'^0, \omega'_{M'})$.  
Let us suppose that this constant
value, $\tilde{q}$, is nonzero.  By the spin-inversion symmetry of
the Hamiltonian (\ref{eq:Hamiltonian}), we must have
\begin{equation}
\label{eq:q0}
\tilde{q}=Q({\cal J},\sigma^0,\omega_M,\sigma'^0,\omega'_{M'})=-Q({\cal
J},\sigma^0,\omega_M,-\sigma'^0,\omega'_{M'})=-\tilde{q}
\end{equation}
for almost every realization.  In the last step we used the fact that
$-\sigma'^0$ can be replaced by $\sigma'^0$ because $Q$ is constant almost
surely (and the distribution of $\sigma'^0$ is spin-inversion symmetric).
It follows from Eq.~(\ref{eq:q0}) that $\tilde{q}=0$.

Now take $M'=M$ and suppose that there 
were a countable number (including the possibility
of a countable infinity) of $M$-spin-flip-stable states.  This would imply
that, with positive probability, two independently chosen starting
configurations and dynamics would result in the same final state, which
would have a self-overlap of $+1$, so that $Q_{M,M}$ would have a delta 
function component at $+1$ with nonzero weight.  It 
follows that, for any finite $M$,
there must be an {\it uncountable\/} infinity of such states. $\, \diamond$

\smallskip

{\it Remark.\/} A crucial step in the proof is the existence of a limiting
final state, i.e., almost sure convergence of the dynamics.  It is the
absence of this knowledge that prevents us from reaching similar
conclusions about ground states (Sec.~\ref{sec:ground}) or pure states at
positive temperature \cite{NS99a} if broken spin-flip symmetry should
exist.  (We note also that in other respects, the method used in this proof
is similar to that used in the proof of Theorem~2 of \cite{NS99a}).  It
follows that the conclusion of Theorem~1 holds also in other models where
the dynamics converge, such as the highly and strongly disordered models
discussed in Sec.~\ref{sec:dynamics} (but in these models the conclusions
can also be obtained by more concrete arguments based on the localization
of the dynamics due to influence nonpercolation, as discussed above).

\smallskip

The conclusion of Theorem~1 (with $M'=M$) holds for a general pair of
$M$-spin-flip stable states.  In Sec.~\ref{sec:samesigma}, we will discuss
how this conclusion is modified for two metastable states dynamically
evolved from a {\it single\/} initial spin configuration.  We discuss now
some of the consequences of Theorem~1, particularly for the proposal that
Hamming distance between metastable states scales with their barrier
height, and that this might lead to a possible ultrametric organization of
metastable states in realistic spin glasses \cite{Orbach1,Orbach2,Orbach3}.
A possible relation of this kind has been conjectured \cite{Nemoto,VV} to
lead to an ultrametric organization of pure states in state space
\cite{MPSTV} in the SK model.

An analysis of this conjecture is hampered by the lack of a clear
understanding of how to define the energy barrier between two metastable
states, 
in the natural context of single-spin-flip Glauber dynamics at
positive temperature.  However, possible progress on these questions has
been made in the mean-field case.  Previous studies \cite{BM80,Evans} have
indicated there that a critical energy $E_c$ exists above which the
(1-spin-flip-stable) metastable states are uncorrelated and have zero
overlap, and below which correlations between barriers and Hamming
distances are expected to emerge.  So it is reasonable to expect that one
should confine one's attention to energetically low-lying metastable states
\cite{VV} (see also the discussion on proper weighting of the states in
\cite{Nemoto}).  It is also clear from general considerations that, because
the distance between two states is symmetric between them but their
relative barriers are not, any analysis should be confined to states with
roughly the same energy (or energy density, in the infinite volume case)
\cite{VV}.

Because we will show in Sec.~\ref{sec:energies} that (for a given $M$ and
choice of dynamical process) almost every metastable state has the same
energy per spin, the above issues are already in part addressed.  But the
more crucial point is that in any subset {\it with nonzero measure\/} of
the set of all metastable states, the same conclusion will hold; namely,
that almost every pair chosen from this subset will have zero overlap. We
conclude that for realistic spin glasses, and supposing that barriers
between states can be defined in some natural way, there should be no
general scaling of barriers with Hamming distance.  This is because almost
every pair of metastable states will have zero overlap, and either almost
every pair also has the same energy barrier or else there's a distribution
of such barriers.  In either case, there's no nontrivial scaling of
barriers with Hamming distance between states.  Furthermore, this
conclusion remains the same when considering metastable states of different
$M$ and $M'$. It should be noted however that these arguments do not rule
out the possibility of some kind of scaling between pairs of states of {\it
zero\/} probability---but such pairs are of negligible significance for
deep quench dynamics \cite{noteoverlaps}.

It might be thought that this conclusion may not apply to finite volumes;
however, we will argue in Sec.~\ref{sec:finitevolume} that the overlap
distribution approximates a delta-function at the origin for large finite
volumes. 

\section{Energies of Metastable States}
\label{sec:energies}

We now turn to a discussion of energies of the metastable states. Our first
result is to show that our dynamical construction yields a probability
measure on the $M$-spin-flip stable configurations such that almost every
one has the same energy density (i.e., energy per site).

\smallskip

{\it Theorem 2.\/} For any dynamical measure $P_M$ (defined as in
Sec.~\ref{sec:dynamics}), almost every $\sigma^\infty_M$ has the same
energy density $E_M$, which is also independent of the coupling realization
${\cal J}$.  

\smallskip

{\it Proof.\/} Because the energy density of any metastable configuration
$\sigma^\infty_M({\cal J},\sigma^0,\omega_M)$ is a measurable,
translation-invariant function of ${\cal J}$, $\sigma^0$, and $\omega_M$,
it immediately follows by the same argument used in Theorem~1 that the
energy density of the $M$-spin-flip stable states is the same for almost
every ${\cal J}$, $\sigma^0$, and $\omega_M$.  $\, \diamond$

\smallskip

The result of Theorem~2 is consistent with the findings of the numerical
investigation of \cite{RO} of the two-dimensional $\pm J$ spin glass, where
the data indicated convergence to a single value of the energy densities of
the (1-spin-flip) metastable states as system size increased.  Although
Theorem~2, as stated, is restricted to systems where the dynamics converge
to a limiting $\sigma^\infty$, which is not the case for the $\pm J$ spin
glass in $2D$ \cite{GNS}, the same arguments imply much more generally
convergence to a {\it single\/} limiting energy density.

Even though our dynamical construction yields a probability measure on the
$M$-spin-flip stable configurations such that almost every one has the same
energy density, it is incorrect to conclude that there does not exist a
spectrum of energy densities among {\it all\/} $M$-spin-flip stable
configurations.  For any $d$ and for most models there will be a nontrivial
spectrum in the sense to be described below; this spectrum can even be
calculated in special circumstances, such as for the $1$-spin-flip stable
states in one dimension \cite{EM}.  (Similarly, although the magnetization
per spin is zero for almost every metastable state, a spectrum of
magnetizations in $1D$ was computed in \cite{DG}.  We will return to this
topic in Sec.~\ref{sec:remanence}.)

To clarify this issue, consider 1-spin-flip stable states in the
continuously disordered spin glass or ferromagnet in $1D$.  The infinite
spin chain can be broken up into ``influence clusters'', as described in
\cite{NS99b,NNS} (see also Subsec.~\ref{subsubsec:hdm}); these are the
finite spin chains bounded to either side by couplings whose magnitudes
satisfy the condition
\begin{equation}
\label{eq:99poundweakling}
|J_{n,n+1}| < \min\{|J_{n-1,n}|,|J_{n+1,n+2}|\} ,
\end{equation}
where the integer $n$ denotes a site along the chain.  Because there is no
frustration, every coupling within every influence cluster is satisfied in
every 1-spin-flip stable state, and the couplings between the influence
clusters --- i.e., those satisfying Eq.~(\ref{eq:99poundweakling}) --- can
be arbitrarily satisfied or unsatisfied.  So one can, for example, take any
percentage $p$ of these ``weak'' bonds to be satisfied and still have a 1-spin-flip
stable configuration, resulting in a spectrum of energy densities among the
set of all 1-spin-flip stable states as $p$ is varied.

This example illustrates the important point that one must be careful in
specifying what measure is imposed on the metastable states before
discussing the distributions of energies, magnetizations, and other
physical quantities over those states.  In the $1D$ example under
discussion, each 1-spin-flip stable state for a given ${\cal J}$ is
specified (modulo a global spin flip) by an infinite sequence of coin
tosses---one for each weak bond.  Here an outcome of ``heads'' on a
particular toss implies that the corresponding weak bond is satisfied, and
``tails'' implies that it is unsatisfied.  The probability measure on the
set of $\sigma_1^\infty({\cal J},\omega_1,\sigma^0)$'s imposed by the
dynamics and initial condition (for fixed ${\cal J}$) corresponds to
independent tosses of an unbiased coin, which is a natural measure for the
purposes of analyzing outcomes of deep quench experiments.  However, one
could arbitrarily impose other measures, for example, those corresponding
to flips of a {\it biased\/} coin; specifically, where the probability $p$
of an outcome of heads on each (independent) flip has $p\ne 1/2$.  For any
such fixed $p$, there are also an uncountable number of 1-spin-flip stable
states (except when $p=0$ or $p=1$), all (outside of a set of measure zero)
with the same energy --- but the energy depends on $p$.

Although this is relatively straightforward for the single-spin-flip case
(even in higher dimension), it becomes more complicated when analyzing
$M$-spin-flip stable states with $M>1$, because now the energies $E_M$ can
in principle depend on the relation between the rates $R_j$ for
$j$-spin-flips (defined in Sec.~\ref{sec:dynamics}) as $j$ varies between
$1$ and $M$.  To see this, consider the case $M=2$, and two different
choices of $P_2$ corresponding to different ratios $R_2/R_1$.  Returning to
the $1D$ chain, consider the final states $\sigma_{2<}^\infty$ and
$\sigma_{2>}^\infty$, gotten when the rates are chosen so that $R_2/R_1\ll
1$ and $R_2/R_1\gg 1$, respectively.  In the former case, the dynamics
allows the system to find (approximately) a 1-spin-flip stable state state
first, in which the probability that any given weak bond is satisfied is
close to $1/2$; the energy associated with the weak bonds is then lowered
further by rigidly flipping neighboring pairs of spins.  In the second case
pairs of spins are flipping rapidly compared to single spins; there is no
reason to expect the energies of $\sigma_{2<}^\infty$ and
$\sigma_{2>}^\infty$ to be the same.  Indeed they can be shown to be
generally different in $1D$ (in the limits $R_2/R_1 \to 0$ and $\infty$) by
a more detailed (but lengthy) analysis.

In general, one might consider for arbitrary $M$ a natural dynamics where,
for each $k<M$, the system first converges to a $k$-spin-flip stable state
$\sigma_k^\infty$ before rigid $k+1$-clusters begin to flip.  Roughly
speaking, this corresponds to a limit where each of the ratios
$R_{k+1}/R_k\to 0$.  A motivation for such a choice is that the systems of
interest are at low temperatures, and a natural scaling with temperature of
the ratio $R_{k+1}/R_k$ is as $\exp[{-f(k)}/T]$, for some positive $f(k)$.
The next theorem is motivated by such a dynamical choice.

For the first part of the theorem, our proof requires more about the common
distribution of the couplings $J_{xy}$ beyond our general assumptions that
the distribution is continuous with a finite mean: namely, that the
possible values of $|J_{xy}|$ include at least three very different
scales---i.e., $J_1, J_2, J_3$ with $J_1/J_2$ and $J_2/J_3$ larger than
some dimension-dependent constant. That will be so (in all dimensions for a
given distribution) if the possible values of $|J_{xy}|$ can be arbitrarily
large or arbitrarily small (or both). This includes Gaussian spin glasses
and disordered ferromagnets (or spin glasses) with a uniform distribution
on $(0,J)$ (or on $(-J,J)$); it does not include disordered ferromagnets
with a distribution on $(J-\epsilon,J)$.

\smallskip

{\it Theorem 3.\/} The energy densities $E_M (R_1,\dots,R_M)$ and $E_{M+1}
(R_1,\dots,R_{M+1})$ satisfy $E_M>E_{M+1}$ providing that $R_{M+1}$ is
sufficiently small for given $R_1,\dots,R_M$ (and the assumption mentioned
above on the coupling distribution is satisfied).  Moreover $E_M$ for any
finite $M$ is larger than the ground state energy density, which (for any
$R_1, R_2,\dots$ ) is the limit of $E_M$ as $M \to \infty$.

\smallskip

{\it Proof.\/} By Theorem~2, for the given $R_1,\dots,R_M$, almost every
$\sigma_M^\infty$ will have the same energy density $E_M$.  For a given
large $t'$, we can choose $R_{M+1}$ small enough so that the energy density
$E_{M+1}(t')$ is as close as we want to $E_M(t')$; this is because
$R_{M+1}$ is so small that only a very tiny density of rigid $M+1$-clusters
have been flipped by time $t'$ (in the $P_{M+1}$ dynamics) so that
$\sigma_M^{t'}$ and $\sigma_{M+1}^{t'}$ are very close.  Furthermore if
$t'$ is large enough, $E_M(t')$ can be made as close as we want to the
limiting value $E_M$. So for any small $\delta$, we can choose first $t'$
and then $R_{M+1}$ so that (a) $|E_{M+1}(t')-E_M| < \delta$ and (b) at most
a density $\delta$ of rigid $M+1$-clusters have been flipped by time $t'$
(in the $P_{M+1}$ dynamics).

The rest of the proof is to show that as time increases from $t'$ to
$\infty$ in the $P_{M+1}$ dynamics, enough other rigid $M+1$-clusters will
flip to lower the energy density from $E_M(t')$ by more than $\delta$. To
do this, it suffices to show that for almost every pair ${\cal J},
\sigma^0$, there exists a density $\rho$ of local configurations (of ${\cal
J}, \sigma^0$) for which one has $M$-spin-flip stability but for which a
flip of some rigid $M+1$-cluster will lower the energy by at least
$\epsilon$. Then the desired result follows by picking $\delta$ small
enough (depending on $\rho$ and $\epsilon$).

Here is one way to find such local configurations.  First, suppose ${\cal
J}$ is such that there is a linear chain of $2M+1$ couplings all of whose
magnitudes, except for the coupling at the very center of the chain, are
very close to some ``large'' value $J_1$.  Suppose further that the center
coupling magnitude is close to an ``intermediate'' value $J_2$ and all
other coupling magnitudes within distance (approximately) $M$ of the linear
chain have magnitudes close to a ``small'' value $J_3$. What is crucial is
not the absolute sizes of the $J_i$'s but that $J_1 \gg J_2 \gg J_3$.  Next
suppose $\sigma^0$ is such that at time zero the $2M$ ``large'' couplings
are all satisfied but the ``intermediate'' center coupling is
unsatisfied. Such a local configuration (which will occur with strictly
positive density because of our assumptions on the coupling distribution)
will have the desired stability properties with $\epsilon$ approximately
equal to $J_2$. Here the rigid $M+1$-cluster to be flipped is half of the
linear chain on either side of the center coupling.

To prove the
final statement, we let $\sigma$ be some ground state and $\sigma'$ be some
$M$-spin flip stable state (with $M$ large). We consider $Z^d$ as the union
of disjoint cubes that are translates of $\Lambda_{L+1}$ with L chosen so that
the volume of each cube is below $M$; each cube should be thought of as an
interior (a translate of $\Lambda_L$) plus boundary. By the metastability,
the restriction of $\sigma'$ to any interior is a finite-volume ground
state for its own boundary condition. Hence if we construct a $\sigma''$
to agree with $\sigma$ on all the interiors and with $\sigma'$ on all
the boundaries, the energy density $E''$ of $\sigma''$ must be higher
than $E'$ of $\sigma'$ (because we no longer have ground states in the
interiors for the boundary conditions). On the other hand, clearly
$E''-E$ is of order $L^{d-1}/L^d$. Thus $E'\le E+O(L^{-1})$ and
hence $\lim_{M\to\infty}E_M \le E$. All ground states have the same 
energy density $E$ (as can be shown by a similar argument) and it
readily follows that the energy density of {\it any} spin configuration
is at least $E$; thus $E_M \ge E$ and 
hence $E_M \to E$ as $M\to\infty$, completing the proof. $\, \diamond$


\section{Remanent magnetization}
\label{sec:remanence}

Suppose that a spin glass with Hamiltonian~(\ref{eq:Hamiltonian}) is
prepared in the uniform initial state $\sigma_x=+1$ for all $x\in Z^d$, and
evolves at zero temperature through the usual 1-spin-flip Glauber dynamics.
What is the typical magnetization of the metastable state into which the
system evolves?  This quantity is of interest because it is related to
experimental measurements of the thermoremanent magnetization in laboratory
spin glasses \cite{FHbook,BY}.  For the continuously disordered spin chain
in one dimension this quantity was found to be $1/3$ \cite{DG,FM}.
Following the practice in those papers we will simply refer to it as the
remanent magnetization and denote it $m_{\rm rem}$.

The question can be recast more generally as finding the value of the
``remanent overlap'' between the initial and final states, $q_{rem} =
\lim_{L\to\infty}|\Lambda_L|^{-1}
\sum_{x\in\Lambda_L}\sigma_x^0\sigma_x^\infty$.  Because of the
translation-invariance of this quantity, it will be constant for almost
every ${\cal J}$, $\sigma^0$ and $\omega_1$; thus no further averaging
(beyond the spatial) is needed. By a simple gauge transformation argument
(see the end of the proof of the next theorem), for a symmetric spin glass
(i.e., where the couplings are symmetrically distributed about zero),
$q_{\rm rem\/} = m_{\rm rem\/}$.  Put another way, the question as to the
value of the remanent overlap is how much direct memory of the initial
state does the final state retain?  This version of the question is as
relevant for random ferromagnets as for spin glasses, and so we will
hereafter address the problem in both its forms: i.e., as the remanent
overlap of a continually disordered system dynamically evolving at zero
temperature from a random initial state, and also as the remanent
magnetization of a symmetric spin glass evolving from a uniform initial
state.

The result of Theorem~2 applies also, as already noted, to the
magnetization per spin, which is zero for almost every $M$-spin-flip stable
state.  However, when the initial state $\sigma^0$ is chosen in a special
way (i.e., all plus), we expect to ``land'' in a 1-spin-flip stable state
with positive magnetization (cf.~the discussion following Theorem~2).  The
next theorem provides a general result for highly disordered models
(Sec.~\ref{sec:convergence}) in any dimension.  We will see that the result
$m_{\rm rem}=1/3$ in the ordinary $1D$ spin glass immediately follows as a
special case.

\smallskip

{\it Theorem 4.\/} Consider the highly disordered model in $d$ dimensions
described in Sec.~\ref{subsubsec:hdm}, undergoing single-spin-flip dynamics
at zero temperature from a random initial state $\sigma^0$.
For almost every coupling realization, $\sigma^0$ and
$\omega_1$, the resulting $\sigma_1^\infty$
will have a remanent overlap with $\sigma^0$ equal to $d/(4d-1)$.
Similarly (and consequently), for a highly disordered symmetric spin glass,
if the inital state is uniform with $\sigma_x=+1$ for all
$x\in Z^d$, then the resulting $\sigma_1^\infty$
will have a remanent magnetization equal to $d/(4d-1)$.

\smallskip

{\it Proof.\/} From the definition of the highly disordered model on $Z^d$,
it follows that any coupling $J_{x_0,y_0}$ that is larger in magnitude than
any of its $2(2d-1)$ neighboring couplings will automatically satisfy the
following condition:  
\begin{equation}
\label{eq:bully}
|J_{x_0,y_0}| > \max\{\sum_{\scriptstyle{z:\|x_0-z\|=1}\atop\scriptstyle{z\ne y_0}}
|J_{x_0,z}|,\sum_{\scriptstyle{z':\|y_0-z'\|=1}\atop\scriptstyle{z'\ne x_0}} |J_{y_0,z'}|\} .
\end{equation}
Therefore, if such a bond is satisfied in
$\sigma^0$, it remains satisfied for all time.  We will refer to these as
``strong'' bonds.  We
will see that these bonds determine the remanent magnetization, so we first
need to compute their density in almost every coupling realization.

The probability of any given bond having this ``strongness'' property is
identical to that of an arbitrary element (call it $X_1$) in a set of
$4d-1$ independent random variables $(X_1,X_2,\ldots,X_{4d-1})$, chosen from a
common continuous distribution, having the highest value in the set.  
(The $X_i$'s here represent the magnitudes of a given coupling and
its $4d-2$ neighboring couplings.) 
Since each $X_i$ is equally likely to be the highest value,
it follows that the probability of an arbitrary
coupling being strong is $1/(4d-1)$.  Then, if $n_s$ is the density
of spins that are located on either end of strong
bonds,
\begin{equation}
\label{eq:remcalchdm}
n_s=\left[1/(4d-1)\right]\times 2\times d = 2d/(4d-1) ,
\end{equation}
where the factor of 2 arises because each strong bond connects to 2 spins,
and the factor of $d$ is the ratio on the lattice $Z^d$ of the number of
bonds to the number of spins.

To find the remanent overlap, we first note that, due to the randomness
of $\sigma^0$, exactly one half of the strong bonds
are satisfied at time zero and will contribute to $q_{\rm rem\/}$, and the other
(unsatisfied) half will not contribute (because in $\sigma^\infty$
every such bond {\it will\/} be satisfied).  What about spins connected
to other bonds?  It was shown in \cite{NN} that the influence clusters of
the strong bonds in the highly disordered model have a tree-like structure,
i.e., contain no loops (this structure on a larger scale, arising for
similar though not identical reasons, also defines the static ground-state
properties of these models; see \cite{NS94,NS96a}).  
Because of the tree-like influence structure, 
the $\sigma_x^\infty$'s  for these other $x$'s are completely
independent of the corresponding $\sigma_x^0$'s
and it follows
that they also contribute zero to the remanent overlap.
Therefore, 
\begin{equation}
\label{eq:remresult}
q_{\rm rem}= \frac 1 2 [2d/(4d-1)] + 0 = d/(4d-1)\, .  
\end{equation}

The last claim of the theorem follows now by a standard gauge transformation
argument, which converts a random $\sigma^0$ into a uniform all plus state, at the
expense of doing a corresponding transformation to the couplings. But for a symmetric
spin glass, the resulting coupling configurations are identically distributed
with the original ones, which completes the proof.  $\, \diamond$

\smallskip

{\it Remark.\/} In one dimension, Eq.~(\ref{eq:remresult}) reduces to
$m_{\rm rem}=1/3$, a result found in \cite{FM} (see also \cite{DG}).  This
is not a coincidence, because the two ingredients used in the proof of
Theorem~3 --- the property that a coupling whose magnitude is greater than
those of any of its neighbors satisfies Eq.~(\ref{eq:bully}), and the
additional property that all influence clusters contain no loops, occur
automatically in any $1D$ model with a continuous coupling distribution.
(In that sense, continuously distributed $1D$ models are already
``highly disordered.'')

\smallskip

What about realistic models in dimensions higher than one?  We present now
a heuristic derivation of a lower bound for $m_{\rm rem}(d)$, based again
on the density of strong bonds.  The condition for a strong bond is given
by Eq.~(\ref{eq:bully}), which involves two independent sums of $2d-1$
random variables, $X_1,\dots,X_{2d-1}$ and $X'_1,\dots,X'_{2d-1}$,
corresponding to the absolute values of the couplings at either end of the
strong bond.  Using the independence of the sums on either side of the
bond, we find the following formula for ${\rm Prob}_d(J_s)$, the
probability of any given bond being strong, where $\tilde{f}_n$ denotes
the probability density function for $X_1+ \dots +X_n$:
\begin{equation}
\label{eq:ferrostrong}
{\rm Prob}_d(J_s) = \int_0^\infty \tilde{f}_1(x)\left\{\int_0^x
\tilde{f}_{2d-1}(y)dy\right\}^2 dx \ .
\end{equation}
Following the same procedure as in Eq.~(\ref{eq:remcalchdm}) yields
a formula analogous to Eq.~(\ref{eq:remresult}),
\begin{equation}
\label{eq:remresult2}
m_{\rm rem} \ge d\,{\rm Prob}_d(J_s)\,.
\end{equation}
However, this expression assumes that the contribution of the spins on all
other bonds is positive or zero.  Although this is plausible, we do not
have a rigorous argument for it, and so the result in
Eq.~(\ref{eq:remresult2}) should be considered heuristic.

The large $d$ behavior of (\ref{eq:ferrostrong}) and (\ref{eq:remresult2})
depends on the nature of the common distribution of the individual
couplings.  For example, if it is a uniform distribution on $[-J,J]$ (so
that $\tilde{f}_1(x) = 1/J$ on [0,J] and zero elsewhere), one finds that
${\rm Prob}_d(J_s)$ behaves as exp$(-4d\log(d) \pm {\rm O}(d))$ as $d \to
\infty$, while for a Gaussian distribution, the $4$ in the exponent is
replaced by $2$. We spare the reader the details of these calculations and
estimates, but note that if on the other hand, the magnitude of the
couplings could neither take on very small nor very large values (e.g., if
the $J_{xy}$'s were uniformly distributed on $[-J,-\epsilon]\cup
[\epsilon,J]$), then ${\rm Prob}_d(J_s)$ would be identically zero
above some dimension.

\subsection{Nature vs.~nurture}
\label{subsec:nvn}

A problem related to remanence is to ask for the extent to which the
final state is determined by the initial spin configuration.  This
should be distinguished from asking for the fraction of spins that have the {\it
same\/} final value as their initial value; rather, we are asking here what
percentage of $\sigma^\infty$ is determined by
$\sigma^0$, where the remainder will depend on the dynamics realization.

In order to quantify this, we introduce a quantity previously considered in
\cite{NNS}.  This quantity, denoted $q_D$, is a kind of dynamical order
parameter somewhat analogous to the Edwards-Anderson order parameter
$q_{EA}$.  Let $\langle \cdot \rangle$ denote the average with respect to
the distribution $P_1$ over dynamical realizations $\omega_1$, for fixed
${\cal J}$ and $\sigma^0$.  We will here use an overbar to denote the
remaining averages over ${\cal J}$ and $\sigma^0$; i.e., with respect to
the joint distribution $P_{{\cal J},\sigma^0} = P_{\cal J} \times
P_{\sigma^0}$.  We then define $q_D = \lim_{t \to \infty}q^t$ (providing
the limit exists, which it does in the ordinary spin glass and random
ferromagnet), where
\begin{equation}
\label{eq:qt}
q^t = \lim_{L\to\infty}(1/|\Lambda_L|)\sum_{x \in \Lambda_L}
\langle \sigma_x^t \rangle^2 = \overline
{\langle \sigma_x^t \rangle^2} \ .
\end{equation}
(The equivalence of the two formulas for $q^t$ follows from
translation-ergodicity.)  When $\sigma^\infty$ exists, then $q_D$ is also
given by the same expressions as in (\ref{eq:qt}) but with $\sigma_x^t$
replaced by $\sigma_x^\infty$.  As already noted, the order parameter $q_D$
measures the extent to which $\sigma^\infty$ is determined by $\sigma^0$
rather than by $\omega_1$ (for fixed ${\cal J}$).  This is because the
middle expression of (\ref{eq:qt}) is the overlap between $\sigma^t$ and
$\sigma'^t$ corresponding to independent replicas $\omega_1$ and
$\omega'_1$ but the same $\sigma^0$ (see also Theorem 6 below).  Of course,
$q^0 = 1$ because $\sigma^0$ is completely determined by $\sigma^0$, while
a value $q_D = 0$ would mean that for every $x$, $\sigma^0$ yields no
information about $\sigma_x^\infty$.  We now present an exact result in one
dimension earlier proved in \cite{NNS}.

\smallskip 

{\it Theorem (Nanda-Newman-Stein) \cite{NNS}.\/}  In the one-dimensional
disordered model with continuous coupling distribution,
$q_D=1/2$.

\smallskip

Because the technical proof appears in \cite{NNS}, we present here only an
informal version.  The idea is that, as discussed earlier, the
one-dimensional chain breaks up into disjoint dynamical ``influence
clusters'', bounded on either end by ``weak'' couplings satisfying
Eq.~(\ref{eq:99poundweakling}).  Each of these
clusters is governed dynamically by a single ``strong'' bond, by which we
mean that once the spin configuration is such that the strong coupling is
satisfied, the state of the spins of $\sigma^\infty$ within
the rest of its influence cluster is completely determined (by the 
signs of the couplings).  Put more
picturesquely, there is a ``cascade of influence'' emanating from the
strong bond and trickling down to either side of its influence cluster
until all couplings within are satisfied.  This means that a spin 
value at $t = \infty$ is
already  determined by (${\cal J}$ and) $\sigma^0$ if the strong bond in
its influence cluster is satisfied at $t=0$, and is completely determined by
$\omega_1$ otherwise.  Because this satisfaction
probability is 1/2, the result
follows.

It is not difficult to extend this result to the highly disordered model
in any dimension \cite{NNS} 
where, because all influence clusters have a tree-like
structure, the idea behind the proof is essentially the same.

For the ordinary spin glass or random ferromagnet, we cannot compute $q_D$
precisely, but it is easy to show that strict inequalities hold at either
end; that is, $0<q_D<1$, so the final state is not
completely determined either by the
initial state ($q_D=1$) or by the dynamics ($q_D=0$).  We refer the reader
to \cite{NNS} (see the proof of Theorem 4 of that paper) for the argument.

\section{Basins of Attraction}
\label{sec:boa}

The basin of attraction of a metastable state $\alpha_M$ may be defined as
the set of starting configurations $\sigma^0$ such that
$\sigma_M^\infty(\sigma^0,\omega_M)=\alpha_M$ for almost every $\omega_M$.
(This generalizes to $M$-spin-flip dynamics the definition given in
\cite{DG}.  A similar definition for the basin of attraction of a {\it
pure\/} state at positive temperature was given in \cite{NS99a}; see also
related discussions in \cite{vEvH}.)  Properties of basins of attraction of
metastable states have played important roles in studies not only of
disordered system dynamics, but also those of neural nets, combinatorial
optimization, and related types of problems where many locally optimal
solutions exist.
Here we ask: how large
(in the sense of the
infinite-temperature (uniform) distribution on spin configurations) 
is the union of all the
domains of attraction of all the metastable states?  

\smallskip

{\it Theorem 5.\/} Under the same assumptions on the coupling distribution as
in Theorem~3, 
almost every initial configuration $\sigma^0$ is
on a boundary between (two or more) metastable states.  Thus, the
union of the domains of attraction of {\it all\/} of the metastable states
forms a set of measure zero (in the space of all $\sigma^0$'s).  

\smallskip

{\it Proof.\/} For $M=1$ (and without the extra assumptions of
Theorem~3), the result follows from the fact stated in
Sec.~\ref{sec:remanence}, and proved in \cite{NNS}, that $q_D<1$ strictly for
disordered models with continuous coupling distributions in any dimension.
That is, for almost every ${\cal J}$ and $\sigma^0$, 
the final state $\sigma_M^\infty$ must depend on the dynamical
realization $\omega_M$; the outcome is not determined purely by
$\sigma^0$.  

To show for $M>1$ that the outcome is not determined purely by $\sigma^0$,
we consider the same type of linear chains of $2(M-1) +1$ couplings
as in the proof of Theorem~3 --- again with all couplings other than 
the center one satisfied at time zero. Then the final state of the
spins along that chain is determined by $\omega_M$, i.e., by which of the
two halves of the chain flips first so that the chain becomes $M$-spin-flip
stable.  $\, \diamond$


\smallskip

{\it Remark.\/} Similar results for pure states were proven in \cite{NS99a}
--- i.e., that, if many pure states exist in the ordinary spin glass in
some dimension $d$ and temperature $T$, then the union of their basins of
attraction form a set of measure zero in the space of all spin
configurations (uniformly distributed in the usual sense).  That a similar
result holds for metastable states is not necessarily surprising, but we
believe that more is true: namely that almost no metastable state lives in
the basin of attraction of a single pure or ground state. This (which we
shall pursue elsewhere) would seem to contradict a standard view in the
literature.

\section{Dynamical evolution from a single initial state}
\label{sec:samesigma}

We now revisit the questions discussed in Sec.~\ref{sec:overlaps} from a
different standpoint. In that section we discussed the nature and
distribution of overlaps for pairs of metastable states (independently)
chosen from the {\it entire\/} set $\{\sigma_M^\infty\}$ of $M$-spin-flip
stable states, generated through our dynamical procedures.  Here we
consider a {\it restricted\/} subset of the 1-spin-flip stable states,
which, although still uncountably infinite, is a set of zero measure of the
1-spin-flip stable states $\{\sigma_1^\infty\}$.  This is the set of states
dynamically generated from a {\it single\/} $\sigma^0$ (chosen from the
usual infinite temperature distribution).  Information on states
chosen from this restricted set may be relevant to studies of damage
spreading \cite{Bag,Grass,JR}, which in some formulations examines overlaps
of pairs of states dynamically generated from the same initial state.

\smallskip

{\it Theorem 6.\/} For fixed ${\cal J}$, consider two metastable states
$\sigma_1^\infty(\sigma^0,\omega_1)$ and
$\sigma_1^\infty(\sigma^0,\omega'_1)$.  To simplify the notation
we will in this section refer to these states as $\sigma^\infty$ and
$\sigma'^\infty$, respectively.  In all cases $\omega_1$ and $\omega'_1$ are chosen
independently.  Then for almost every such pair, the spin overlap
equals $q_D>0$, where $q_D$ is the dynamical order parameter defined as
the $t\to\infty$ limit of $q^t$ in Eq.~(\ref{eq:qt}).

\smallskip

{\it Proof.\/} Throughout this proof we suppress the dependence of the two
final states on $\sigma^0$, because both metastable states are understood
to evolve from the same initial state; we also suppress the $M=1$
subscript on $\sigma^\infty$.  Then the overlap of the two final
states is  
\begin{equation}
\label{eq:sameoverlap}
\lim_{L\to\infty}|\Lambda_L|^{-1}\sum_{x\in\Lambda_L}\sigma_x^\infty(\omega_1)
\sigma^\infty_x(\omega'_1)=E_{{\cal J},\sigma^0,\omega_1,\omega'_1}
\left[\sigma_x^\infty(\omega_1)\sigma^\infty_x(\omega'_1)\right]\, ,
\end{equation}
where $E_{\cal J}$ denotes an average with respect to the distribution over
the couplings, and similarly for the other distributions.
Eq.~(\ref{eq:sameoverlap}) follows from the translation-ergodicity of the
distributions from which the couplings, initial state, and dynamical
realizations are chosen, along with the translation-invariance of the
overlap.  Because the dynamical realizations $\omega_1$ and $\omega'_1$ are
chosen independently, it follows that
\begin{equation}
\label{eq:ind}
E_{{\cal
J},\sigma^0,\omega_1,\omega'_1}\left[\sigma_x^\infty(\omega_1)\sigma^\infty_x(\omega'_1)\right]
= E_{{\cal J},
\sigma^0}\left[E_{\omega_1}\left(\sigma_x^\infty(\omega_1)\right)\right]^2=q_D\ . \, \, \, \diamond
\end{equation}

\section{Finite volumes}
\label{sec:finitevolume}

Most of the preceding discussion concerns infinite-volume disordered
systems.  Because experiments and numerical simulations are done on finite
systems (and in the latter case often not very large ones), it is important
to study how the theory of metastable states constructed so far is modified
when attention is restricted to finite volumes.  It has often been the case
for conventional homogeneous systems that both thermodynamic and dynamical
behavior in infinite systems is a straightforward extrapolation from
behavior in large finite volumes; but recent work has shown that for
disordered systems such simple extrapolations can often fail, and in
general the relationship between the physics of finite and infinite systems
can be subtle \cite{NS97,NS96b,NS98}.

We therefore re-examine many of the questions previously raised and
answered for infinite systems in the context of a finite system on a cube
$\Lambda_L$ of volume $V=L^d$ spins.  The first question we will address is
how the number of metastable states scales with volume.  We showed in
Theorem~1 that for infinite systems the number of $M$-spin-flip metastable
states is uncountably infinite in any dimension; it is natural then to
expect that this number scales exponentially with volume (in a
$d$-dependent fashion) for finite systems, and we will now prove that this
this is true in general.  In some models, like the ordinary $1D$ disordered
chain, or the highly disordered model in general dimensions, the scaling
behavior of the number of 1-spin-flip stable states can be calculated
exactly.

\smallskip

{\it Theorem 7.\/} Let $N_{M,d}(V)$ denote the number 
of $M$-spin-flip stable states in the cube $\Lambda_L$ of volume $V=L^d$
in $d$ dimensions.  Under the same assumptions on the coupling 
distribution as in Theorem~3: for the ordinary spin 
glass and random ferromagnet
$N_{M,d}(V)=\exp\left[O(V)\right]$, in the sense that $N_{M,d}(V)$
is bounded above by $\exp\left[a_M^+(d)V\right]$ and below by
$\exp\left[a_M^-(d)V\right]$, where the coefficients $a_M^-(d)>0$
and $a_M^+(d) < \infty$
depend on the model chosen.  In both the highly disordered spin glass and
highly disordered random ferromagnet, for $M=1$, $V^{-1} \ln \left[N_{1,d}(V)
\right]$ converges to $a_1(d)=(d\ln 2)/(4d-1)$ as $V \to \infty$.

\smallskip

{\it Remark.\/} As already mentioned, results for the highly disordered
model apply also to the ordinary $1D$ disordered chain, where the
coefficient becomes $a_1(1)=(\ln 2)/3$, in agreement with earlier
calculations \cite{Li,DG}.

\smallskip

{\it Proof.\/} We will prove the second claim first.  Computation of the
exact number of 1-spin-flip stable states in the highly disordered model
consists of two parts: computing the density of strong bonds (satisfying
Eq.~(\ref{eq:bully})), and showing that the number of 1-spin-flip stable
states corresponds to the number of ways to satisfy all the strong bonds.

The density of strong bonds in the highly disordered model was already
computed in the proof of Theorem~4, in the discussion preceding
Eq.~(\ref{eq:remcalchdm}).  From that discussion, the average number of
strong bonds $n_b(d,V)$ in volume $V$ satisfies
\begin{equation}
\label{eq:strongavg}
V^{-1} n_b(d,V) \to \left[d/(4d-1)\right] \, .
\end{equation}
Each strong bond (which must be satisfied in all $M$-spin-flip stable
states, for any $M$) can be satisfied in two ways, corresponding to a 
simultaneous flip of the two spins at either end of the bond.  To complete
this part of the argument, we need to show that the number of
1-spin-flip stable states equals $2^{n_b(d,V)}$.

To do this, we note that the 1-spin-flip dynamics breaks $Z^d$ up into
disjoint influence clusters, as shown in \cite{NNS,NN}. These have a
tree-like structure, so that under 1-spin-flip dynamics each has two
possible spin configurations, related by a global spin flip; the spin
configuration of each influence cluster is determined entirely by that on
the strong bond.  

To prove the first claim of Theorem~7, we will establish lower and upper
bounds for $N_{M,d}(V)$ in the ordinary spin glass or disordered
ferromagnet.  A trivial upper bound is obtained by noting that the number
of metastable states cannot exceed the total number of spin configurations,
so that for any $M$, $a_M^+(d)\le\ln 2$.  To establish an $M$-dependent
lower bound, we consider first the case $M=1$.  The density of strong bonds
(obeying Eq.~(\ref{eq:bully})) was computed in Eq.~(\ref{eq:ferrostrong}),
but it is sufficient for our purposes here to note simply that under the
assumptions of Theorem~3 on the coupling distribution, this density is
positive is any dimension.

Even though influence percolation may occur in these models, the strong
bonds are still satisfied or unsatisfied independently of one another (and
once satisfied, remain so for all time), as in the highly disordered case.
Thus there are (approximately) $2^{d[{\rm Prob}(J_s)]V}$ ways for the
strong bonds to be satisfied and at least an equal number of 1-spin-flip
stable states.

The proof is completed by noting that for $M>1$, we may consider the
same type of linear chains of $2(M-1)+1$ couplings as in the proof of
Theorem~3. Here the center couplings of the chains play the role
of the strong bonds and the density of center couplings replaces
${\rm Prob}(J_s)$ in obtaining a lower bound for $a_M^-(d)$.
This completes the
proof.  $\, \diamond$

\smallskip

We now turn to the important question of whether the results obtained so
far for infinite systems --- in particular, the answers 4) -- 6) discussed in
Sec.~\ref{subsubsec:main} --- hold (to an increasingly good approximation
as system size increases) in large finite volumes.  The answer to 2),
showing convergence of the dynamics (within a volume $\Lambda_L$ and with
specified boundary conditions) to a limiting $\sigma_{(L)}^\infty$ was already
provided in Sec.~\ref{subsec:finiteconvergence}. 

Why is it important to study this?  The reason is that it is not
necessarily the case {\it a priori\/} that the answers to 4) -- 6) would
hold, even roughly and in a qualitative sense, for large finite volumes in
the limit as $L\to\infty$.  It could conceivably be the case, for example,
that the overlaps between final states evolved from two arbitrarily chosen
initial states, and with independent dynamics, might {\it not\/} be
concentrated about zero in finite volumes of arbitrarily large size (even
though the overlap would be exactly zero for the infinite volume system
according to Theorem~1 of Sec.~\ref{sec:overlaps}); instead, it might be,
if one looked at many pairs of initial states and dynamical realizations,
that one would find a distribution of final state overlaps spread over many
values, which would {\it not\/} increasingly concentrate about zero as
$L\to\infty$.  This would be a type of dynamical analogue to the
``nonstandard SK picture'', or a similar thermodynamic scenario, raised as
a logical possibility in \cite{NS97,NS96b} (but ruled out as a viable
option through a combination of rigorous and heuristic arguments in
\cite{NS98}).

We will now show that such scenarios should not in fact occur; that is, the
answers to 4) -- 6) will hold to a good approximation in large finite
volumes, and with increasing accuracy as their size increases.  So, to use
the example in the preceding paragraph, we would find that the distribution
of overlaps between final states evolved from pairs of arbitrarily chosen
initial states, and with independent dynamics, would be clustered about
zero in an increasingly tight distribution as $L\to\infty$.  We will prove
this rigorously for both highly and strongly disordered models (the latter
of which has similar thermodynamic behavior to an ordinary spin glass or
random ferromagnet), and will provide convincing heuristic evidence that
the same remains true for ordinary disordered models. Our strategy will be
to show that the final state $\sigma_{(L)}^\infty$ agrees with the
infinite-volume $\sigma^\infty$ (in a way to be made precise momentarily)
increasingly well as $L\to\infty$.  (As always, our results are for almost
every state; in the finite-volume context, this means the exclusion of an
increasingly small probability event, typically exponentially small, in the
volume.)

We now make these ideas more precise.  Consider a volume $\Lambda_L$ with
specified boundary conditions, such as free, fixed, or periodic.  As always
we take $\Lambda_L$ to be a $d$-dimensional cube of side $L$ centered at
the origin.  Consider within that cube a smaller one, denoted
$\Lambda_{L'}$, also centered at the origin, and with $L'<<L$.  The
boundary conditions on $\Lambda_{L'}$ may be the same as those on
$\Lambda_L$ or different.  Consider now the two states $\sigma_{(L)}^t$ and
$\sigma_{(L')}^t$, generated from a pair of initial states and a pair of
dynamical realizations that, in each case, are identical within the smaller
volume $\Lambda_{L'}$.  We define the ``region of agreement'' (at time $t$)
between $\sigma_{(L)}^t$ and $\sigma_{(L')}^t$ as the set of sites $x$
within $\Lambda_{L'}$ where $\sigma_{(L)x}^t=\sigma_{(L')x}^t$.

We want to ask whether (for most initial states and dynamical realizations)
the fraction of sites in $\Lambda_{L'}$ belonging to the region of
agreement at time $t = \infty$ is close to one.  More
precisely, we want to know whether if we take first the limit $L\to\infty$
and then $t\to\infty$, the agreement fraction approaches one
as $L' \to \infty$.  If so,
then we would be finished.

Let us examine this in more detail.  Consider, for example, periodic
boundary conditions on both $\Lambda_L$ and $\Lambda_{L'}$.  Because a
limiting final state exists in each volume, the probability that the spin
$\sigma_x$ at any particular site $x$ has not reached its final state,
i.e., will flip again, after a time $\tau_x$, must go to zero as $\tau_x$
increases for fixed $L'$ and $L$.  
If this probability goes to zero independently of $L'$ and $L$ as both
become large (i.e., if the probability $g_{L'}(\tau_x)$ in $\Lambda_{L'}$
is bounded by an $L'$-independent function $g(\tau_x)$ that goes to
zero), then we're done.  Put another way, eventually (as system
sizes increase) the effects of the receding boundaries (even as $t\to\infty$)
are felt
increasingly less.

To see why this proves the result, we can use this probability to choose a
time $\tau$ where, say, 95\% of the sites in $\Lambda_{L'}$ have reached
their final configuration, and this time is independent of $L'$.  Now
compare this to the restriction to $\Lambda_{L'}$ of the corresponding
infinite-volume $\sigma^\infty$.  After the time $\tau$, the only spins
within $\Lambda_{L'}$ that ``notice'' they're subject to periodic boundary
conditions would be those within some distance of order one (as
$L' \to \infty$) of the
boundary.  The others reach the same state as in the infinite system, and
so the overlaps agree in that region.

This argument clearly will hold when $M=1$
in any model where influence percolation
does not occur, such as highly or strongly disordered models.  In those
systems, all dynamics is localized, as discussed in Sec.~\ref{subsubsec:hdm}.
Therefore, as $L\to\infty$, there will be some $L$ beyond which every spin
in $L'$ will reach the same state as in the infinite system; that is, every
influence cluster will be unable to distinguish (dynamically) whether it
belongs to a finite or infinite system.  Although the argument was
presented in an informal way, this is sufficient to prove the result,
stated formally as Theorem~8.

\smallskip

{\it Theorem 8.\/} For single-spin-flip dynamics 
in any model where influence percolation does not occur,
such as ordinary $1D$ disordered chains, or both the highly and strongly
disordered models, the distributions of overlaps, 
energies \cite{finitemean}, and other global
properties of metastable states in large finite volumes approaches the
infinite-volume results as the volumes tend to infinity.

\smallskip

While this argument is rigorous when $M=1$
for models without influence percolation,
it does not carry over easily to $M>1$ (except for $1D$ where a modified
influence percolation argument can be carried out) or to
ordinary disordered models in dimensions
greater than one.  Heuristically, though, the same result should
apply there too.  In order for it not to do so, it would have to be the
case that the final energy density in finite volumes, for some specified
boundary conditions, would be {\it lower\/} (by an amount {\it not\/}
tending to zero with volume) than that in the infinite
system.  But the absence of boundary conditions in the infinite system
means that, in any finite subvolume, the spin configuration can dynamically
adjust to the fixed coupling realization at the boundaries in order to
attain the lowest possible energy; it is difficult to see why the energy
should be lower when this option is not available due to the boundary
condition being rigidly imposed externally, and without regard to the
couplings.

But even if this were so, it would still be irrelevant to the state
observed on any numerically or experimentally accessible timescale.  This
is because, in the infinite volume case, the system relaxes to a final
state within a finite subvolume in some finite time.  This same time would
set the scale for an initial relaxation of a large finite-volume system.  There
must then be an additional timescale, depending on $L$, for information
generated at the boundary to propagate to spins deep in the interior,
changing their state.  This new timescale must diverge as $L\to\infty$
because of the finite signal propagation time imposed by the dynamics
(Sec.~\ref{sec:dynamics}); that is, for large enough volumes the region of
agreement of the final states generated by finite-volume and
infinite-volume dynamics would be most of $\Lambda_L$, up to timescales
diverging with $L$.

The scenario described in the last paragraph is unlikely, however, because
it is already unlikely that finite-volume energy densities are lower
than those for infinite-volume systems.  It is noted only to show that, for
any practical scenario of experimental interest, the results of Theorem~8
should hold also for $M>1$ and for 
ordinary disordered systems in any finite dimension.

\section{Ground states}
\label{sec:ground}

All of our preceding discussion has concerned metastable states, stable up
to $M$-spin flips.  These are generated by a dynamics with distribution
$P_M$, in which lattice animals up to size $M$ are rigidly flipped as
described in Sec.~\ref{sec:dynamics}.  It is natural to ask what happens if
we let $M\to\infty$; in particular, can a dynamics that allows rigid flips
of lattice animals of unbounded size be constructed so as to generate
infinite-volume ground states?  We will address that question in this
section and see that the answer (when 
formulated carefully) is yes.  However,
unlike the case for finite $M$, we cannot show convergence to a final
state (and indeed, convergence may not be valid, as we discuss below), 
and so cannot obtain results of the kind generated for metastable
states.  We will also discuss several issues related to the connection
between ground states and $M$-spin-flip stable states, in both finite and
infinite volumes.

We therefore consider the ``lattice animal dynamics'' introduced
in Sec.~\ref{sec:dynamics}, now with the lattice animal size unbounded.
The rates $R_k$ were already chosen 
so that the dynamics, as specified by clock rates
(or equivalently, mean waiting times for a given lattice animal to
attempt to flip), ensures that information doesn't propagate infinitely
far in a finite time and so there is a well-defined dynamics
(see Sec.~\ref{sec:dynamics}).
The assumptions on the $R_k$'s imply the following lemma,
which will be needed to prove the next theorem.

\smallskip

{\it Lemma.\/}  Consider a volume $\Lambda_L$ and a given lattice
animal $A$ that is entirely inside $\Lambda_L$ (i.e., no spins in $A$
touch the boundary $\partial\Lambda_L$).  Then at an
arbitrarily chosen time $t$, the probability $p_1$ that the clock 
of $A$ ``rings''
(i.e., it attempts to flip) before time $t+1$ and before the clock
of {\it any\/} other lattice animal
touching $\Lambda_L$ or $\partial\Lambda_L$,
is strictly positive (independently of $t$ or the spin configuration
at time $t$).

\smallskip

{\it Proof.\/} This follows immediately from the nature of the dynamics
(whose distribution is denoted hereafter by $P_\infty$) because
of our assumptions on
the rates $R_k$ needed to make the process well-defined.
In particular, if we denote the number of sites in a lattice
animal $A$ by $|A|$ and denote by $R^{(L)}$ the (finite) sum
of $R_{|B|}$ over all lattice animals $B$ that touch $\Lambda_L$
or its boundary, then 
\begin{equation}
\label{eq:prob1}
p_1 = (R_{|A|}/R^{(L)})\,(1-e^{-R^{(L)}})\,.
\end{equation}
%
\smallskip

We now show that the dynamics defined by $P_\infty$ leads to a ground
state, in the sense to be discussed below.

\smallskip

{\it Theorem 9.\/} Consider the dynamics with distribution $P_\infty$, and
a finite $\Lambda_L$ of arbitrary size.
Then after a random time $t_L$ (depending on $L$, ${\cal J}$, $\sigma^0$,
and dynamics realization $\omega_\infty$), the spin configuration inside
$\Lambda_L$ forever remains in a ground state subject to its boundary
conditions (where the ground state and the boundary
condition could themselves change with time).

\smallskip

{\it Proof.\/} We note first that,
as always, (with probability one)
any fixed lattice animal can undergo only finitely many
energy-lowering flips.  This then implies that the following event must
have zero probability: there exists an infinite
sequence of times $t_1, t_2, \ldots
\to\infty$ such that at each of those times, the spin configuration
inside the cube (given its boundary conditions at that time) is not in a
ground state configuration.  This is because, after any of those times, the
above Lemma implies that there is a positive probability in the next unit
of time that some lattice animal strictly {\it inside\/} the cube flips to
lower the energy.  The finiteness of $L$ implies a finite number of lattice
animals inside $\Lambda_L$, so that if this event did not have zero
probability, then, with positive probability,
some lattice animal inside the volume would flip
infinitely many times.  $\, \diamond$

\smallskip

We emphasize a few points, most importantly, that there is no claim that
the dynamics converges to a specific
ground state $\sigma^\infty$ (though it might,
depending on dimension and disorder distribution).  The proof of
convergence for finite $M$ (\cite{NNS}
and Sec.~\ref{subsec:infiniteconvergence} above)
fails here because now the energy per
spin of a lattice animal flip of size $M$
can go to zero as $M\to\infty$.  Of course, if
convergence to a ground state $\sigma^\infty$ can be shown for a
particular model,
it would immediately imply (cf.~Theorem~1) that there would be an
uncountable number of {\it ground\/} states, and their overlap distribution
function would be a delta-function at the origin (see also discussions in
\cite{NS97,NS98}).  It is therefore of interest to pursue this question,
but we will not do so here.

 
A second point is that our dynamics ``algorithm'' finds ground states in
the sense that any finite region surrounding the origin will 
eventually always be in
some ground state (no energy-lowering flips possible within the region)
after some time (depending on the various realizations as discussed
in Theorem~9).  It could still happen, though, that spins within the region
flip infinitely often (as they must if there are {\it not\/} uncountably
many ground states, as is expected, e.g., in the $2D$ random ferromagnet).
These could occur either through a rigid flip of the entire region, or
through changes in boundary conditions due to flips of large lattice animals
intersecting the region.

Finally, we note that this is a rare example of a dynamical process that
can be proved to lead to a Gibbs state (in this case, a ground state at
$T=0$).  While it is widely expected that finite-temperature Glauber
dynamics, and similar dynamics that satisfy detailed balance, lead to Gibbs
states at positive temperature, as $t \to \infty$, we are unaware of any 
general proof (for a discussion of related $T>0$ results, see Sec.~IV.5 of \cite{Liggett}).

It may seem surprising that there can be an uncountable number of states
energetically stable to rigid flips of $M$ spins, where $M$ can be
arbitrarily large (but fixed), and yet there exists only a single pair of ground
states.  Yet this is precisely what happens in disordered $1D$ chains, and
almost certainly as well in the $2D$ disordered ferromagnet.  (Recent
numerical evidence also points towards only a single pair of ground states
in the $2D$ spin glass as well \cite{Middleton,PY}.)  Caution should
therefore be exercised whenever information on ground (or pure) states is
used to extract information on metastable states, or vice-versa.

\section{Conclusions}
\label{sec:conclusions}

We began in Sec.~\ref{subsubsec:questions} with a list of ten questions about
basic properties of metastable states in disordered systems, providing
brief answers in Sec.~\ref{subsubsec:main} followed by a detailed study
in subsequent sections.  
These questions and answers aimed towards understanding fundamental
features of the set of $M$-spin-flip stable states in
spin glasses and disordered ferromagnets, such as their numbers, basins of attraction,
energies, overlaps, remanent magnetizations, and relations to thermodynamic states.

From a broader perspective, we have presented a viewpoint for considering
metastable states in spin glasses and random ferromagnets; its essence is that one can construct a systematic approach
towards their study, just as has been traditionally done for fundamental statistical
mechanical objects such as spin configurations or thermodynamic states.  We approach the
problem of metastability as in those cases, by noting that one is often most interested
(with exceptions as discussed) in the typical states that appear in a physically relevant
ensemble for the particular problem under study.  In the case of spin configurations,
this enemble is usually the Gibbs state at a given temperature; in the case
of thermodynamic states, we have proposed in previous papers \cite{NS97,NS96b,NS98} that
the appropriate ensemble is the metastate. In the current context of metastable
states, we propose a natural ensemble (on the $\{\sigma^\infty(\sigma^0,\omega_M)\}$'s)
that arises from zero-temperature ``lattice animal''
dynamics evolving from a spin configuration generated through a deep quench; we
call this $M$-dependent measure the $M$-stable ensemble.

To summarize, we propose the following comparison:

\bigskip

\begin{tabular}{ccc}
$\underline{\rm Object}$& &$\underline{\rm Ensemble}$\\
Spin configuration&$\Longrightarrow$&Gibbs ensemble\\
Gibbs state&$\Longrightarrow$&Metastate ensemble\\
Metastable configuration&$\Longrightarrow$&$M$-stable ensemble
\end{tabular}

\bigskip

We suggest that this dynamical approach provides both a natural
ensemble and the corresponding tools for studying metastable states.

\medskip

\renewcommand{\baselinestretch}{1.0}


\begin{thebibliography}{10}
\bibitem{noteoverlaps} For example, one might consider the following
situation.  Allow single-spin-flip dynamics to run until convergence to a
specific $\sigma_1^\infty$.  Using that as the starting state, raise the
temperature to a small nonzero value $T$ and allow positive-temperature
Glauber dynamics to run for a specified time $\tau$, after which the
temperature is reduced back to zero.  It is not unreasonable to speculate
that the Hamming distance of the final state $\sigma'^\infty_1$ reached in
this way from the initial $\sigma_1^\infty$ increases with $\tau$, for
$\tau$ small.  Such considerations may be relevant to some aging
experiments \cite{aging1}, but our analysis makes clear that any possible
correlations of this kind cannot be extended to provide information on the
properties of the larger set of metastable states.  This example
illustrates that one needs to be careful to specify what is meant, for
example, by correlations between barriers and overlaps; the meanings may be
determined by the experimental situation under investigation.
\bibitem{finitemean}
The energy density results are for strongly disordered coupling
distributions with finite mean.
\end{thebibliography}
\end{document}